% ---- Tex file -------------
\baselineskip=12pt
\magnification=\magstep1
\def\ref{\par\noindent\hangindent=0.3cm\hangafter=1}
\def\lsim{\mathrel{\rlap{\lower 3pt\hbox{$\mathchar"218$}}
     \raise 2.0pt\hbox{$\mathchar"13C$}}}
\def\gsim{\mathrel{\rlap{\lower 3pt\hbox{$\mathchar"218$}}
     \raise 2.0pt\hbox{$\mathchar"13E$}}}

\centerline { Revised Version (30  July 1996)}
\bigskip
\bigskip

\centerline {\bf THE X-RAY BACKGROUND AS A PROBE} 
\centerline {\bf OF DENSITY FLUCTUATIONS AT HIGH REDSHIFT} 

\bigskip

\centerline{ Ofer Lahav$^1$, Tsvi Piran$^2$ and Marie A. Treyer$^1$ }

\bigskip

(1) Institute of Astronomy, Madingley Road,  Cambridge CB3 0HA, UK 

(2) Racah Institute  of Physics, The Hebrew University, 
Jerusalem 91904 , Israel

\bigskip
\bigskip

\noindent{\bf Abstract:} 
The X-Ray Background (XRB) probes structure on scales intermediate
between those explored by local galaxy redshift surveys and by the
COBE Microwave Background measurements.  We predict the large scale
angular fluctuations in the XRB, expressed in terms of spherical
harmonics for a range of 
assumed power-spectra and evolution scenarios. 
The dipole is due to large scale structure as well as  to the
observer's motion (the Compton-Getting effect).  For a typical
observer the two effects turn out to be  comparable in amplitude. 
The coupling of the two effects makes it difficult 
to use the XRB for independent confirmation
of the CMB dipole being due to the observer's motion.
The large scale structure dipole 
(rms per component) relative to the 
monopole is in the range   
$a_{1m}/a_{00} \sim (0.5-9.0) \times 10^{-3} $.
The spread is mainly due to the assumed 
redshift evolution scenarios  
of the X-ray volume emissivity $\rho_x(z)$.
The dipole's prediction is consistent with a  measured dipole in the
HEAO1 XRB map.
%For example, assuming  $\Omega=1$ with a power-spectrum  fitting  local
%galaxy surveys (and rms normalizatio n $\sigma_8=1$),
%  on 8 $h^{-1}$ Mpc spheres),   
%and X-ray volume emissivity $\rho_x(z) \propto (1+z)^2$ out
%to redshift $z_{max}=5$, the predicted quadrupole (rms per component)
%relative to the XRB mean intensity is $a_{2m}/{\bar I} \approx 4
% \times 10^{-3}$.  Similar predictions are also given for the standard
% CDM model and a range of evolution models.  
Typically, the harmonic
spectrum drops with $l$ like $a_{lm} \sim l^{-0.4}$.  This behaviour
allows us to discriminate a true clustering signal against the flux
shot noise, which is constant with $l$,  
and may dominate the signal unless 
bright resolved sources are
removed from the XRB map.  We also show that
Sachs-Wolfe and Doppler (due to the motion of the sources) 
effects in the XRB are negligible.  
Although our analysis focuses on the XRB,  the formalism is general 
and can be easily applied to other cosmological backgrounds.

\bigskip

{\bf 1. Introduction}   
\bigskip

Although discovered before the Cosmic Microwave Background (CMB), the origin of
the X-ray Background (XRB) is still unknown.  
But it seems  likely
that the XRB is due to sources at high redshift 
(for reviews see Boldt 1987; Fabian \& Barcons 1992).
Here we shall not attempt to speculate on the nature of the XRB sources.
Instead, we {\it utilise} the XRB as a probe of the density fluctuations at
high redshift.  The XRB sources are probably
located at redshift $z < 5$, making them convenient tracers of the mass
distribution on scales intermediate between those in the CMB as probed
by COBE ($\sim 1000 $ Mpc), and those probed by optical and IRAS redshift
surveys ($\sim 100 $ Mpc).  In terms of the level of anisotropy, the XRB is
also intermediate between the tiny CMB fluctuations ($\sim 10^{-5}$ on angular
scales of degrees) and galaxy density fluctuations (of the order of unity on
scale of 8 $h^{-1}$ Mpc). 

In recent years 
the
XRB has been studied by means of 
analysing the total intensity, 
the spectrum and the spatial fluctuations.
In particular, the spatial fluctuations were analysed by:
(i)  Source identifications of high-flux regions 
(e.g. Shanks et al. 1991);
(ii) Auto-correlation functions  for which upper limits 
 and marginal detections were reported (e.g. de Zotti et al. 1990, 
Jahoda \& Mushotsky 1991,
Carrera et al. 1993, Chen et al. 1994, So\l tan \&  Hasinger 1994);
(iii)  Cross-correlation of the XRB with  galaxies and clusters for 
which the detections and interpretation are reasonably established 
(e.g. Lahav et al. 1993, Miyaji et al. 1994, 
Carrera et al. 1995, Barcons et al. 1995, Roche et al. 1995, 
So\l tan et al. 1996, 
Treyer \& Lahav 1996).

The preliminary measurements of the dipole anisotropy 
in the XRB (Shafer 1983; Shafer \& Fabian 1983, Boldt 1987)
were  discussed qualitatively
by associating it
with local clusters such as Virgo and the Great Attractor and by other
cosmographical arguments (e.g. Rees 1979; Fabian \& Warwick 1979;
Warwick, Pye \& Fabian 1980;
Jahoda \& Mushotzky 1989; Goicoechea \& Martin-Mirones 1990).  In this
paper we treat the problem in a statistical rather than cosmographical way.
We generalize the analysis for any spherical
harmonic of order $l$, corresponding to angular resolution $\theta
\sim \pi/l$.  The predicted rms harmonics  are derived in
the framework of growth of structure by gravitational instability from
density fluctuations drawn from a Gaussian random field.  The
harmonics are then expressed in terms of the power-spectrum of density
fluctuations and for evolution scenarios which are consistent with
recent measurements of galaxy clustering and the Cosmic Microwave
Background.  As there is quite a lot of freedom in the
parameterization of the XRB sources we shall restrict ourselves in
this paper to an Einstein-de Sitter universe ($\Omega=1$,
$\lambda=0$), 
although 
some of the expressions evaluated below are
also valid for other  world models. 
The Hubble constant is given as $H_0
= 100 h$ km/sec/Mpc.  In principle, the X-ray background(s) should be
discussed in  different frequency bands, e.g. in the hard band (2-10
keV, e.g. HEAO1) and in the soft band (0.5-2.0 keV, e.g. ROSAT), which
exhibit different properties.  However, the current uncertainty in
measurements (e.g. Table 1 in Treyer \& Lahav 1996) does not make it
practical at present to distinguish between the different bands.  The
formalism is kept general and can be used for specific cases in the
future.  It can also be easily generalized to other cosmological 
backgrounds.

The outline of the paper is as follows. Section 2 presents the
harmonic formalism and rms predictions. 
Our main result is given in equations 15 and 16 and the reader who
is not interested in the mathematical details can skip directly to 
these equations.
Numerical estimates based on these formulae are
given in Section 3, and a comparison to the observed HEAO1 XRB dipole is
discussed in Section 4. 
We discuss the results in Section 5.  
In Appendix  A we show that the Sachs-Wolfe and Doppler effects for the XRB  
are negligible compared with the source density fluctuations.

\bigskip
\bigskip

{\bf 2. Spherical Harmonic Expansion of Background Sources} 
\bigskip

We consider a cosmological population of XRB sources that
trace the matter distribution and examine the angular
fluctuations in the observed XRB surface brightness. 
For convenient comparison 
with the  Cosmic Microwave Background (CMB) 
(e.g. Padmanabhan 1993)
and with 
galaxy distributions at low redshift 
(e.g. Peebles 1973, Scharf et al. 1992, Fisher, Scharf \& Lahav 1994)
we expand 
the surface brightness of the XRB over the sky in  spherical
harmonics and we estimate the expected rms fluctuations of different
multipoles. 
For large scales (low multipoles)
these fluctuations  might be larger than the Poisson noise, 
provided bright resolved sources are removed from the XRB map.

\bigskip
\bigskip

{\it 2.1 The Clustering Term}
\bigskip
The XRB surface brightness 
 $I_{\nu_0} ({\bf \hat r}) d{\nu_0}$
is observed in  a {\it narrow} 
frequency band $(\nu_0, \nu_0 + d \nu_0)$.
Hereafter we  omit the 
frequency label (whenever it is not essential) 
to make the notation easier. We
expand $I({\bf \hat r})$ in  spherical
harmonics
\footnote{$^{\star}$}{We consider here an
ideal detector with zero beam width. If the detector has a beam profile 
of size $\theta_B$, then harmonics of order $l \sim \pi/\theta_B$
are washed out.
Since we  are  mostly
interested in lower harmonics this effect is negligible for our
calculations.}: 
$$
I({\bf \hat r}) = \sum_{lm}  a_{lm}  Y_{lm}({\bf \hat r}) .
\eqno (1) 
$$

The XRB most likely results from numerous discrete sources.
In this case 
the harmonic coefficients, $a_{lm}$, 
can be derived by summing over the sources, 
each with observed flux $f_i (\nu_0) $.
$$
a_{lm} = \sum_{sources} f_i (\nu_0) \; Y_{lm}^*({\bf \hat r_i}) .
\eqno (2)
$$
The flux observed in the frequency  band ($\nu_0, \nu_0 + d \nu_0) $ 
due to an individual source emitting at frequency $\nu = \nu_0 (1+z) $ 
at redshift $z$ and luminosity distance $r_L$ is :
$$
f ({\nu_0}) d {\nu_0}  = { L_{\nu}  \over { 4 \pi r_L^2} } d \nu = 
{ L_{\nu_0 (1+z) }  \over { 4 \pi r_L^2} } (1+z) d \nu_0 .
\eqno (3)
$$
Note the extra $(1+z)$ factor which is due to the observation being 
per unit frequency. The predicted harmonic coefficients are then:
$$
a_{lm} = \int \int dV_c  \; dL_{\nu} { { L_{\nu} (1+z)}
 \over { 4 \pi r_L^2} }\; [1 + b_x \delta(r_c, {\bf \hat r})] 
\; \Phi(L_{\nu}, z) \; Y_{lm}^*({\bf \hat r}),
\eqno (4)
$$ 
where again $\nu = \nu_0 (1+z)$. 
Here $\delta$ is the mass density perturbation,
and we have assumed that the {\it comoving} 
luminosity function $\Phi(L_{\nu}, z)$
is independent of local overdensity. 
For an $\Omega=1$ universe the volume element is 
$dV_c=r_c^2 dr_c^2 d\omega$,
where 
$r_c ={ 2c \over H_0} [1 - (1+z)^{-1/2} ]$ 
and the luminosity distance is $r_L = (1+z) r_c$.
We have also assumed that there is a linear biasing
between the X-ray sources and the mass fluctuations: 
$$
\delta_x(r_c,\hat r) = b_x \; \delta(r_c,\hat r),
\eqno (5)
$$
for all redshifts. This is of course a naive assumption, reflecting
our poor knowledge of the way X-ray sources are formed with cosmic time
relative to the mass fluctuations.

The observed comoving luminosity density (`volume emissivity')
due to sources at redshift $z$ is:
$$
\rho_x(z) =  \int dL_{\nu} \; L_{\nu}\;  \Phi(L_{\nu}, z) (1+z) ,  
\eqno (6)
$$
where again $\nu = \nu_0 (1+z)$.
% For example, if the  spectral energy distribution of the sources is
% $I_{\nu} \propto \nu^{-\alpha}$, and their space density  
% evolves like $(1+z)^p$,
% then we can write : 
Herafter we assume a simple power-law evolution model for 
the X-ray light density:
$$
\rho_x (z) =  \rho_{x0} (1+z)^{q}.
\eqno (7)
$$
For example if 
the spectral energy distribution of a source
is $L_{\nu} \propto \nu^{-\alpha} $
and the source density evolves like $(1+z)^p$ then
$q = p - \alpha + 1$.

For the monopole ($l=0$), with $Y_{00} = (4 \pi)^{-1/2}$,  
 we recover the 'Olbers integral' or Lookback factor 
(cf. e.g. Weinberg 1972, Boldt 1987).  
In the case $\Omega=1$, $\Lambda=0$,
the mean intensity 
%(in units of ergs/sec/cm$^2$/str)
out to redshift $z_{max}$ is:
$$
{\bar I} =  
{a_{00} \over {\sqrt {4 \pi}    } } = 
\int dV_c {  {\rho_x(z)} \over { 4 \pi r_L^2} }    
= { \rho_{x0} \over 4 \pi} ({ c \over H_0}) 
\int_0^{z_{max}} dz (1+z)^{q -7/2}
\eqno (8)
$$ 
The {\it fluctuations}  in the background are expressed 
by  harmonics $l > 0 $ :
$$
a_{lm} = {1 \over {4 \pi} }   \rho_{x0}  { c \over H_0} 
\int \int d \omega\;  
dz   (1+z)^{q -7/2} \;
b_x \delta ( r_c, {\bf \hat r})\; 
Y_{lm}^*({\bf \hat r}) \;.
\eqno (9)
$$

It is convenient to expand  the density contrast  
in Fourier modes (where $k$ is in comoving coordinates) :
$$ 
\delta ({\bf r_c}) \equiv 
{ {\delta \rho} \over \rho } ({\bf r_c}) =   
{1 \over ({2 \pi}) ^3 } \int d^3k \; \delta_{\bf k} \;
 e^{- i {\bf k}\cdot {\bf r_c}}\;,  
\eqno (10)
$$
and to use 
the Rayleigh expansion of a plane wave in spherical coordinates :
$$
e^{i {\bf k}\cdot {\bf r_c}} = 4 \pi \sum_{lm} i^l j_l(kr_c) 
Y_{lm}^* ({\bf \hat r_c}) 
Y_{lm} ({\bf \hat k})\;. 
\eqno (11)
$$
With the orthogonality condition
$\int d \omega Y_{lm} ({\bf \hat r}) Y_{l'm'}^* ({\bf \hat r}) = 
\delta_{ll'}^{mm'}$,
we get:
$$
a_{lm} = (i^l)^* { 1 \over { 2 \pi^2} } {1 \over {4 \pi}}
\rho_{x0}  { c \over H_0}  
\int d^3 k \; b_x \delta_{\bf k} (z) Y_{lm}^* ({\bf \hat k} ) 
\int dz (1+z)^{q-7/2}  j_l(kr_c), 
\eqno (12)
$$
where $j_l$ is the Bessel function of order $l$.
It is convenient to parameterize the growth of density perturbations by: 
$$
\delta_{\bf k}(z) = \delta_{\bf k}(0) \;  (1+z)^{-\mu}.
\eqno (13)
$$
E.g. in linear theory in an Einstein-de Sitter universe $\mu=1$, 
which is a reasonable parameterization  for the low-order harmonics
(i.e. the large scales).
The power-spectrum is given in terms
of the present day fluctuations  $\delta_{\bf k}$ as:
$$ 
\langle \delta_{\bf k} \; \delta_{\bf k'}^*  \rangle \; =  
(2 \pi)^3  P(k) \delta^{(3)}({\bf k} - {\bf k'} )\; . 
\eqno (14)
$$
Taking  the mean-square values and using Parseval's theorem 
we obtain the prediction for the rms fluctuations
per harmonic component (there are $2 l + 1$ 
components per $l$ and as the model is isotropic they are all equal 
and independent of $m$):
$$
\langle |a_{lm}|^2 \rangle  = { 1 \over (2 \pi)^3} \; 
\rho_{x0}^2 b_x^2 ( { c \over H_0} )^2 
\int dk k^2 P(k) |\Psi_l(k)|^2 \;.
\eqno (15)
$$   
%As discussed in Appendix C, 
Only fluctuations within the horizon grow.
Hence unless the initial power spectrum is very peculiar most of
the contribution to the integral in Eq. (15) is 
from short wavelengths. In this case 
assuming the specific evolution model of eqs. (7) and (13), 
we can write
the window function as:
$$
\Psi_l(k) \approx
\int_{z_{min}}^{z_{max}} 
 dz (1+z)^{q - \mu -7/2}  j_l(k r_c)\;, 
\eqno (16) 
$$

In principle, 
there could be a Sachs-Wolfe (1967) contribution to the XRB harmonics 
due to the difference in potential between 
the sources and the observer (similar to the effect in the CMB fluctuations
on scales $> 10^o$), 
and a Doppler contribution due to the 
motions of the XRB sources and to our motion. 
However, as shown in Appendix A, 
the Sachs-Wolfe  and Doppler 
(due to  the XRB sources motion)  effects 
are less than  $\sim 0.1 \%$ 
of the clustering effect and can safely be ignored.
\
\bigskip
\bigskip
{\it 2.2 The Shot Noise Term}
\bigskip

On the other hand, a significant signal may arise from  
shot noise, due to the discreteness of the objects: 
$$
\langle |a_{lm}|^2 \rangle_{sn}  = {1 \over 4 \pi} \sum_{sources}  f_i^2\;. 
\eqno (17) 
$$   
A flux cutoff $f_{m}$  must be  used
to eliminate bright sources, 
to avoid divergence of eq. (17).
  In terms of the differential
number-flux relation in  Euclidean space, $N(f) = N_0 f^{-2.5}$, the
shot noise is: 
$$
\langle |a_{lm}|^2 \rangle_{sn} = \int_0^{f_{m}} f^2 N(f) df 
= 2 N_0 f_{m}^{0.5} \; \;  \propto r_{m}^{-1},    
\eqno (18)
$$
where $r_{m} = \sqrt { L_*/ 4 \pi f_{m} } $ is 
the effective  cutoff distance for an $L_*$ galaxy.    
We give estimates of the shot noise relative to the clustering  
signal at the end of the next section.
The shot-noise term is constant with $l$ unlike the clustering term,
which allows us, at least in principle, to distinguish between the
two.

\bigskip 
\bigskip

{\bf 3. Quantitative Predictions for the  XRB} 
\bigskip

To visualize the scales probed by the XRB,  we show in Figure 1 the product 
$ k^3 P(k) \sim \langle ({{\delta \rho } \over \rho})^2 \rangle $ 
for  the standard Cold Dark Matter (CDM) power-spectrum $P(k)$ with
$\Gamma \equiv  \Omega h = 0.5$ (Bardeen et al 1986) 
and for  a low density CDM (LDCDM) power-spectrum with $\Gamma = 0.2$.
We regard the LDCDM  power-spectrum only  as a phenomenological 
fit to the clustering of local  galaxies 
(e.g.  Fisher et al. 1993),
and  we retain $\Omega=1$ in the rest of the  analysis. 
The window functions $|\Psi_2(k)|^2$ 
are shown for the quadrupole   
($l=2$) of:
(i) the IRAS 1.2Jy redshift survey,  
(ii) the XRB (eq. 16 with  
$q - \mu=3 $,  $z_{min}=0$,  $z_{max}=5$) 
and (iii) the CMB, where$^*$ $|\Psi_2(k)|^2 = k^{-4}j_2^2(2ck/H_0)$.
\footnote{}{\hskip -.7cm $^*$The CMB harmonics due to the 
Sachs-Wolfe effect in $\Omega=1$ universe are (e.g. Padmanabhan 1993):
$$
<|a_{lm}|^2 >_{SW} = \big({ H_0 \over 2c}\big)^4 \; { 2 \over \pi} 
\int dk k^{-2} P(k) [j_l(2ck/H_0)]^2 \;.
%\eqno (19) 
$$}  
We see that  the XRB  indeed probes intermediate scales,
filling in the `gap' between local redshift surveys and COBE.

Figure 2 shows the prediction for the XRB rms harmonics 
(per $\{l,m\}$ component) 
due to source clustering (eq. 15).
We normalize the rms $a_{lm}$ by the monopole 
$a_{00} = {\sqrt {4 \pi} }  { \bar I} $ (eq. 8),  
% $\rho_{x0} (c/H_0) (b_x \sigma_8 )$,  
so that the ratio is dimensionless, and we do not have to specify 
$\rho_{x0}$. We also divide by the unknown normalization 
$b_x \sigma_8$, the present-epoch rms fluctuation of X-ray sources 
in $8 h^{-1}$ Mpc sphere,
which is probably of the order unity.
The  mass density fluctuation
$\sigma_8$ 
can be specified e.g. from the COBE CMB measurements 
(for standard CDM $\sigma_8 = 1.35$ ; Sugiyama 1995).
The normalization $b_x \sigma_8$ can also be fixed by 
comparing the measurement of the XRB auto-correlation function
(e.g. So\l tan \& Hasinger) with the prediction in terms of the 
power-spectrum (e.g. Treyer \& Lahav 1996), 
but this determination is beyond the 
scope of this paper.

The normalized harmonics are shown in Figure 2 for both 
standard and low density CDM models 
(both with density perturbation growth index $\mu=1$, 
$z_{min} =0$, and $z_{max}=5$),
for a rather extreme   evolution 
parameter $q=4$.
The harmonics decline monotonically like $a_l \propto l^{-0.4}$.
Values for the normalized dipole ($a_{1m}/a_{00}$) 
and quadrupole ($a_{2m}/a_{00}$)
are given in Table 1 for $q=4$ and $q=0$ (no evolution).
We see that the predictions are very sensitive to $q$, 
i.e. to the redshift evolution of $\rho_x(z)$.
They are relatively little dependent on 
the assumed
power-spectrum  and on the maximal redshift $z_{max}$.

The observational test for the detection of a clustering 
signal can be obtained 
by looking for a monotonically declining signal with $l$, compared 
with a constant shot-noise term.
For the hard-band (2-10 keV) we can estimate the shot noise (eq. 18) 
by adopting 
$N_0  \approx 2 \times 10^{-15}$  
(erg/sec cm$^{-2}$)$^{3/2}$ str$^{-1}$ 
and $f_{m} = 3 \times 10^{-11}$ erg/sec  cm$^{-2}$, 
above which sources were identified
(Piccinotti et al. 1982). 
With the observed  mean intensity  (Boldt 1987) 
${\bar I} = a_{00}/{ \sqrt {4 \pi} } =
 5.2 \times 10^{-8}$ ergs/sec/cm$^2$/str 
we find that  the shot-noise normalized  
to the monopole is 
$
\langle |a_{lm}|^2 \rangle_{sn}^{1/2}/a_{00} \approx
 8.0 \times 10^{-4}.
$
This shot-noise level is 
comparable to  the predicted dipole and quadrupole in $q=4$ models
but well below the expected LSS signal for $q=0$ model (see Table 1). 
Therefore the signal-to-noise ratio strongly depends 
on the redshift evolution of $\rho_x(z)$.  
It is important therefore to explore a range of models
and procedures for shot-noise suppression.
In particular, there is some  freedom to choose $f_{m}$ such that 
the signal to noise is maximized.

We illustrate this point by 
applying a similar calculation to the soft-band
using the observed 
ROSAT source counts (Georgantopoulos et al.~1996 ) extrapolated to unresolved 
fluxes. 
We find that for the rather extreme evolution model $q=4$ 
the first 10 multipoles outreach the shot noise if sources 
brighter than $f_{m}\approx 10^{-14}{\rm erg~cm^{-2}s^{-1}}$ are removed.
Note that for consistency they must also be removed from the 
$a_{lm}$'s,
hence reducing the clustering signal.
Calculating the multipoles as a function
of flux limit requires knowledge of the X-ray luminosity
as a function of redshift. For our purposes, we simply apply a
redshift (radius) cutoff to Eq. (16). 
We find that the low order multipoles
still outreach the shot noise (assuming the above flux cutoff) 
if sources nearer than  few 100  h$^{-1}$ Mpc 
are removed from the X-ray map, although the harmonic spectrum flattens
significantly.
Again, we emphasize that 
for other evolution models (e.g. the extreme no-evolution 
case $q=0$)  the expected clustering signal is well above 
the shot-noise.

\bigskip
\bigskip

{\bf 4. XRB Dipole } 
\bigskip

The dipole pattern in the XRB is due to two effects: (i) the flux
emitted by the XRB sources tracing the large scale structure (LSS), as
we predicted above (eq. 15 for $l=1$); and (ii) the motion of the
observer relative to the XRB.  The second effect, first discussed by
Compton \& Getting (1935) for the cosmic-ray background, gives a
dipole pattern of the form
$$
{ \Delta I \over {\bar I} } 
= (3 + \alpha) { V_{obs} \over c } \cos (\theta),
\eqno (19)
$$ 
for an observer moving at velocity $V_{obs}$ relative to an isotropic
sea of radiation with spectrum $I_\nu \propto \nu^{-\alpha}$.  This
relation is most easily derived from Liouville's theorem, which
implies that $I_{\nu}/{\nu}^3$ is an invariant.  This was accurately
measured in the CMB (where $\alpha=-2$), most recently using the COBE
4-year data (Lineweaver et al 1996), giving a solar motion of $368.9
\pm 2.5$ km/sec relative to the CMB in the direction $(l=264^o;
b=48^o)$.  
 However, based on this measurement alone,  we cannot rule
out an entropy gradient origin for the CMB dipole (Pac\'ynski and Piran 
1990;  Langlois and Piran 1996). 
If  the CMB dipole does arise from 
our motion relative to the CMB background,
then we expect to find a similar
contribution to the XRB dipole. 
For the hard XRB, with $\alpha=0.4$, the expected excess in the
direction of motion is  
$ { \Delta I \over {\bar I} } = 4.2 \times 10^{-3} $. 
But as we show below,  it is unfortunately 
difficult to separate the Compton-Getting (CG) effect
from the dipole due to large scale structure (LSS) in the distribution of the 
XRB sources, as the two effects have similar amplitudes.

The measurements of the XRB dipole are not accurate, due to 
contamination by Galactic
emission and low resolution, but several studies reported a detection.
The HEAO1 2-10 keV whole-sky map shows a dipole (Shafer 1983, Shafer
\& Fabian 1983, Boldt 1987) in the direction $(l=282^o; b=30^o)$ (the
90\% confidence region is rather large and covers about 1/8 of the
sky).  If the entire signal is due to motion, then the inferred
velocity is $475 \pm 165$ km/sec.  At higher energies (80-165 keV) the
dipole's direction is $(l=304^o; b=26^o)$ and the derived velocity
(again assuming the dipole is purely due to motion) is
$1450 \pm 440 $ km/sec (Gruber 1991).  It is perhaps not too surprising
that (within the large error bars) the derived velocity is larger than
that deduced from the CMB, as the XRB dipole may be `contaminated' by
 LSS anisotropies.  The importance of the LSS effect due to
nearby unresolved sources is supported by several studies.  Jahoda \&
Mushotzky (1989) found an enhancement in the direction of the Great
Attractor (see also Goicoecha \& Martin-Mirones 1990), Miyaji \& Boldt
(1990) derived from a sample of AGNs an acceleration dipole which is
consistent with the CMB dipole's direction, and a cross-correlation
signal was detected between the unresolved XRB and nearby galaxy
catalogues (e.g. Lahav et al 1993, Miyaji et al. 1994, Carrera et
al. 1995).

Our formalism allows us to estimate the strength of the 
two effects for a hypothetical random observer.
As shown in Table 1
the expected LSS  dipole  is in the range  
$a_{1m,LSS} / a_{00} \sim  (0.5-9.0) \times  10^{-3}$. 
To estimate the Compton-Getting (CG) effect
we first calculate the mean square 
velocity for a random observer
(e.g. Kaiser \& Lahav 1989, Padmanabhan 1993): 
$$
\langle V_{rms}^2 \rangle ={ {H_0^2 \Omega_0^{1.2}} \over {2 \pi^2} } 
\int dk \; P(K)  e^{-k^2 R_*^2},
\eqno (20) 
$$
assuming linear theory and
that the density fluctuations causing the motion are at distances
much smaller than the horizon.
The region sharing the motion is modeled here as a Gaussian sphere
with radius $R_*$.  
For a point ($R_*=0$) typically 
$V_{rms} \approx 1000  \sigma_8 \Omega^{0.6} $ for the CDM models concerned. 
\footnote{$^{\star}$}
{When compared with e.g. the motion of the Local Group relative to the CMB 
($\approx 600$  km/sec), one should take 
a filtering scale $R_*$ of a few Mpc, leading to lower predicted 
rms velocities.
}
The 
expected CG dipole in the rms sense is then 
$a_{1m,CG}/a_{00}  = { 1 \over 3 }  (3+\alpha) 
{ V_{rms} \over c } $.
%=  3.7 \times 10^{-3}    { V_{rms} \over { 1000 {\rm km/sec} } } $.
As shown in Table 1 
it  is interesting  that the CG and LSS are of comparable
amplitude for a hypothetical observer, 
including the case of 
our Sun's motion.

It is important to verify that the XRB CG dipole agrees with 
the CMB dipole as a proof that the CMB dipole is due to motion.
However, as the `contamination' by the LSS effect  
is unknown, 
it is better to subtract  the CG 
dipole (based on the CMB dipole) from the XRB map,
and to look at the residual LSS effect.
Jahoda (1993)  removed the CG dipole from the HEAO1 (2-10 keV) map, 
and after correcting for Galactic emission (according to Iwan et al. 1982), 
found  
$|D_{LSS}| \equiv |\sum f_i {\bf \hat r}_i|  \approx 3 \times 10^{-9}$
erg/sec/cm$^2$ towards $(l=309^o; b=+45^o)$. 
%In our notation, and taking 
%${ \bar I}  = 5.2 \times 10^{-8}$ ergs/sec/cm$^2$/str
%in  that band, 
%the residual LSS dipole amplitude 
%  $a_{1m,LSS}/ {\bar I} = D/( \sqrt{4 \pi}  {\bar I})
%  \approx 0.016$.
In our notation 
$a_{1m,LSS}/ a_{00}  = D/( {4 \pi}  {\bar I})$.
We see from Table 1 that the observed residual LSS dipole 
is within the range of our model predictions 
for the LSS dipole. 
A more detailed estimation of the HEAO1 dipole is underway  (Scharf et al., 
in preparation).

\bigskip
\bigskip

{\bf 5. Discussion} 
\bigskip

This paper gives quantitative predictions for 
the fluctuations on large angular scales in the X-ray Background.
The rms  predictions are based on
assumed power-spectrum and evolution scenarios, 
and are expressed in spherical harmonics.
We stress that any application to  whole-sky XRB maps
(such as HEAO1 and ROSAT) requires careful treatment 
of the shot-noise (by removing bright sources) and
the smearing by the beam size.
Another major observational obstacle (in particular for
estimating the quadrupole in the soft band) is Galactic emission, 
but it can be corrected for by inversion and filtering techniques.

Our main conclusions are:

(i) The XRB is an important  probe
of density fluctuations on scales intermediate
between scales explored by galaxy redshift surveys and  by 
COBE.  

(ii) For a range of cosmological and evolution models 
the shape of the harmonic
spectrum drops with $l$ like $a_{lm} \sim l^{-0.4}$.
The amplitude of the harmonics is mainly sensitive to the 
resdshift evolution  the X-ray volume emissivity 
$\rho_x(z) = \rho_{x0}\;  (1+z)^{q}$.  
We show that for some models (e.g. $q=4$) the signal is comparable to 
the shot-noise, while for others(e.g. $q=0$), 
the signal-to-noise ratio is $\sim  10$. 
 
The assumed power-spectra and maximal redshift little affect the predictions.

(iii) For realistic models, the harmonic amplitudes for $l <  10$ 
are above the shot-noise level (which is constant with $l$), 
provided that bright resolved sources are
removed from the XRB map.  

(iv) Sachs-Wolfe and Doppler effects in the XRB are negligible
compared to the clustering signal.

(v) The expected dipole amplitude due to large scale
 structure is comparable to the Compton-Getting dipole amplitude due
 to the observer's motion, 
and is consistent 
with a recently measured dipole in the HEAO1 XRB map.
Unfortunately, the coupling of the two effects makes it difficult 
to detect the CG dipole in the XRB independently
(unless the LSS dipole amplitude is 
predicted from a model or by extrapolation
from higher harmonics).

As illustrated in this paper,
important cosmological information can therefore be obtained by analysing 
whole-sky XRB maps, in particular 
the HEAO1 (2-10 keV) and ROSAT (0.5-2.0) surveys.

\bigskip

\noindent{\bf Acknowledgments:}
We thank A. Baleisis, E. Boldt, A. Fabian, K. Fisher, K. Jahoda, 
R. Kneissl, D. Langlois,
T. Miyaji, C. Scharf and N. Sugiyama for helpful discussions, and the
referee for suggesting to improve the shot-noise calculation.  OL and
MAT acknowledge the hospitality of the Hebrew University in Jerusalem,
and in particular that of A. Dekel.  MAT is grateful to the European
Community for a postdoctoral fellowship.

\bigskip

\noindent {\bf References}
\bigskip
 
\ref  Barcons X., Franceschini A., de Zotti G, 
	Danese L., Miyaji T., 1995, ApJ 455, 480. 
\ref Bardeen, J.M., Bond, J.R., Kaisr, N., \& Szalay,A., 1986,
            ApJ, 304, 15  
\ref Boldt E., 1987, Phys. Reports, 146, 215
\ref Carrera F.J., Barcons X., Butcher J.A., Fabian A.C., 
  	Stewart G.C., Toffolatti L., Warwick R.S., Hayashida K., 
	Inoue H., Kondo H., 1993, MNRAS, 260, 376
\ref Carrera F.J., Barcons X., Butcher J.A., Fabian A.C., 
  	Lahav O., Stewart G.C., Warwick R.S., 1995, ApJ, 275, 22
\ref Chen L.-W., Fabian A.C., Warwick R.S., Branduardi-Raymont G.,
  	Barber C.R., 1994, MNRAS, 266, 846 
% \ref Cole, Treyer \& Silk (1992), ApJ, 385, 9
\ref Compton, A. \& Getting, I., 1935, Phys Rev, 47, 817
% \ref Efstathiou et al 199?;
\ref Fabian, A.C. \& Warwick, R.S., 1979, Nature, 280, 39
\ref Fabian A.C., Barcons X., 1992, ARA\&A, 30, 429
\ref Fisher, K.B., Davis, M., Strauss, M.A., Yahil, A., Huchra, J.P., 
       1993, ApJ, 402, 42
\ref Fisher, K.B., Scharf, C.A. \& Lahav, O., 1994, MNRAS, 266, 219
\ref Georgantopoulos, I., Stewart,  G.C., Shanks, T., Griffiths R.E.,
      \& Boyle, B.J., 1996, MNRAS, 280, 276
\ref Goicoecha L.J. \& Martin-Mirones J.M., 1990, MNRAS 244, 493
\ref Gruber D.E., 1991, 
{The X-ray Background}, X. Barcons \& A. Fabian
	Eds., Cambridge University Press.
%\ref Hasinger G., Burg R., Giacconi R., Hartner G., Schmidt M., 
%	Tr\" umper J., \& Zamorani G., 1993, A\&A 275, 1.
\ref Iwan, D. et al. 1982, ApJ, 260, 111
\ref Jahoda, K.,  1993, Adv. Space Res., 13, (12)231
\ref Jahoda K. \& Mushotzky R.F., 1989, ApJ, 346, 638
\ref Jahoda K. \& Mushotzky R.F., 1991,
{The X-ray Background}, X. Barcons \& A. Fabian
	Eds., Cambridge University Press.
\ref Kaiser, N., Lahav, O., 1989, MNRAS, 237, 129
\ref Langlois, D.,   Piran, T.,  1996, Phys Rev D, in press
\ref Lahav O., Fabian A.C., Barcons X., Boldt E., Butcher J., Carrera
  	F.J., Jahoda K., Miyaji T., Stewart G.C., Warwick R.S., 1993,
  	Nature, 364, 693
% \ref Lahav et al. 1994
\ref  Lineweaver, C.H., Tenorio, L.,  Smoot, G.F., 
        Keegstra, P., Banday A.J., \& Lubin, P., 1996, ApJ Lett, submitted
\ref Miyaji, T. \& Boldt, E., 1990, ApJ Lett, 353, L3 
\ref Miyaji, T., Lahav O., Jahoda K., Boldt E., 1994, ApJ, 434, 424
\ref Pac\'ynski, B.,  Piran, T., 1990, ApJ, 364, 341 
\ref Padmanabhan, 1993, {\it Structure Formation in the Universe}, 
Cambridge University Press, Cambridge.
\ref Peebles, P.J.E., 1973, ApJ, 185, 413
%\ref Peebles P.J.E., 1980, {\it The Large Scale Structure of 
%	the Universe}, Princeton University Press, Princeton
\ref Piccinotti, G., Mushotzky, R.F., Boldt, E.A., Holt, S.S., 
Marshall, F.E., Serlemitsos, P.J., Shafer, R.A., 1982, ApJ, 253, 485
\ref Rees M., 1979, {\it Objects of High Redshifts}, IAU Symp.,
	G. Abell \& J. Peebles Eds.
\ref Roche N., Shanks T., Georgantoulos I., Stewart G.C.,
	Boyle B.J., Griffiths R., 1995, MNRAS, 273, L15
\ref Sachs R.K. \&  Wolfe A.,  1967, ApJ, 147, 73
\ref Scharf, C.A., Hoffman, Y.,  Lahav, O., \& Lynden-Bell, D., 1992 
       MNRAS, 256, 229
\ref Shafer, R.A., 1983, PhD Thesis,  University of Maryland, NASA TM-85029 
\ref Shafer, R.A. \& Fabian, A.C., 1983, {\it Early Evolution of
	the Universe and its Present Structure}, IAU Symp. No. 104,
	p. 333, G.O. Abell \& G. Chincarini Eds.
\ref Shanks T., Georgantopoulos I., Stewart G.C., Pounds K.A., 
	Boyle B.J., Griffiths R.E., 1991, Nature, 353, 315
\ref So\l tan A. \& Hasinger G. 1994, A\&A, 288, 77
\ref So\l tan A., Hasinger G., Egger R., Snowden S., Tr\"umper J.,
	1996, A\&A 305, 17
\ref Sugiyama, N., 1995, ApJ Supp, 100, 281
\ref Treyer M.A. \& Lahav O., 1996, MNRAS,  280, 469
\ref Warwick R.S., Pye J.P. \& Fabian A.C., 1980, MNRAS 190, 243
\ref Weinberg S., 1972, {\it Gravitation and Cosmology: Principles
	and Applications of the General Theory of Relativity}, 
	J. Wiley \& sons, NY
\ref de Zotti G. et al., 1990, ApJ, 351, 22

\bigskip
\bigskip

\centerline{\bf Appendix  A: The Sachs-Wolfe and Doppler effects
for the X-ray Background}
\bigskip

\def\bfr{{\bf r}}

\def\bfk{{\bf k}}

\noindent The fluctuations in the gravitational potential 
and velocity field
at the time
the XRB is produced yield  additional variations.
Ignoring non-linear effects, 
if the fluctuations are small they are additive and  we can modify 
 Eq. (9) to be (for $l >0$):
$$
a_{lm} = {1 \over  4\pi}  
\rho_{x0} { c \over H_0} 
\int \int d\omega  dz  (1+z)^{q - 7/2}
\left[ b_x \delta + (3+\alpha)\left(
 { {\delta \phi} \over {3 c^2} } +   
 {( {\bf V}  - {\bf V}_{obs}) \over c}\cdot {\hat {\bf r} }\right) \right] 
 \; Y_{lm}^*({\bf \hat r}),
\eqno (A1)
$$ 
where $\delta \phi$ is the perturbation in the potential, 
${\bf V} $ is the peculiar velocity of the XRB sources, 
${\bf V}_{obs} $ is the velocity of the observer in the same
reference frame as ${\bf V}$ and $\alpha$ is the 
spectral energy distribution index.
The factor  $(3+ \alpha)$ was inserted according to the Compton-Getting 
formula (eq. 19) for the observer's motion.
The same factor applies to the motion of the XRB sources (Doppler) and 
to the
gravitational redshift  (Sachs-Wolfe) 
effect as they are all redshift effects.
The  division of the effect between Sachs-Wolfe and Doppler which we have
written here is in the synchronous gauge. This division is gauge
dependent but the total result is not (Padmanabhan, 1993).

We shall assume $\Omega=1$ and linear theory. 
We expand $\delta$ and $\delta \phi$
in Fourier series and relate them via Poisson equation:
$$
\delta \phi_{\bf k} = {3 \over 2} { H^2 a^2 \over k^2 }  \delta_{\bf k}(z)
={3 \over 2} {H_0^2}  (1+z) { \delta_{\bf k}(z)  \over k^2}\;.
\eqno (A2)
$$
The last equality follows as 
$(Ha)^2 =H_0^2 (1+z)$ in an Einstein de-Sitter universe.

The line-of-sight velocity can be written in terms of spherical harmonics
and $\delta_{\bf k} $  
by decomposing  the 
potential  into Fourier components
and using 
Rayleigh's expansion  and linear theory 
(e.g. Fisher, Scharf \& Lahav 1994):
$$
U(\bfr) 
= {{ H a} \over{2\pi^2}}\sum\limits_{lm} (i^l)^* \int d^3k\, 
{{\delta_{\bfk}}(z) \over k}
j_l^{'}(kr) Y_{lm}^*(\hat\bfk)Y_{lm}(\hat\bfr)
\eqno (A3)
$$
where $j_l^{'}(kr)= dj_l(kr)/d(kr)$ is
the first derivative of the Bessel function.

Following the  analysis in Section 2 , with 
$\delta_{\bf k} (z) = \delta_{\bf k}(0) \;  (1+z)^{-1}$ 
and  $\rho_x(z) = \rho_{x0}(1+z)^{q}$, 
we can write Eq. (A1) as:
$$
a_{lm} = (i^l)^* { 1 \over { 2 \pi^2} } {{\rho_{xo}} \over {4 \pi}} 
\int d^3 k  Y_{lm}^* ({\bf \hat k})\delta_{\bf k} 
\left[ 
b_x { c \over H_0} \Psi_{l,\delta} + 
{ (3+\alpha) \over 2 k^2}  { H_0 \over c}  \psi_{l,SW} + 
 { (3+\alpha) \over k} \psi_{l,D}
\right]
$$
$$ 
-{ 4 \pi \over 3} (3+\alpha)  {  {|{ \bf V}_{obs}|} \over  c}  
~{\bar I}~ Y^*_{lm}({\hat {\bf v}}_{obs} ) ~\delta_{l1}
\eqno (A4)
$$
The last term is  due to the motion 
of the observer (cf. Compton-Getting eq. 19), 
where 
${\bar I}$ is the mean intensity (eq. 8), 
and ${\hat {\bf v}}_{obs}$ is the direction of motion of the observer.

The window functions (under the assumption that fluctuations 
only grow within the horizon) are:
%as explained  in Appendix C)
$$
\Psi_{l,\delta} (k) \equiv 
\int_{z_{min}}^{z_{max}}  dz  \;  (1+z)^{q - 9/2}\;  j_l(k r_c)\;, 
\eqno (A5)
$$
$$
\Psi_{l,SW} (k) \equiv 
\int_{z_{min}}^{z_{max}}  dz  \;  (1+z)^{q - 7/2}\;  j_l(k r_c)\;, 
\eqno (A6)
$$
and 
$$
\Psi_{l,D} (k) \equiv 
\int_{z_{min}}^{z_{max}}  dz  \;  (1+z)^{q - 4}\;  j_l^{'}(k r_c)\;. 
\eqno (A7)
$$
Finally, taking the mean square value of (A4) we obtain, for the first three 
terms:
$$
< |a_{lm}|^2 > = { \rho_{x0}^2 \over (2 \pi)^3}  
\int dk k^2 P(k) 
\left[ 
b_x { c \over H_0} \Psi_{l,\delta} + 
{(3+\alpha)\over 2k^2}  { H_0 \over c} \psi_{l,SW} + 
 { (3+\alpha) \over k} \psi_{l,D}
\right]^2
\eqno (A8)
$$ 
The interpretation of the 3 terms that multiply 
$k^3 P(k) \sim \langle ({ \delta \rho \over \rho })^2 \rangle$ 
can be understood as follows.
Apart from the $k$-dependence of the window functions, 
the first term squared is constant with $k$, 
the second term (Sachs-Wolfe) squared  scales 
like $k^{-4}$ and the third term (Doppler) squared 
scales like $k^{-2}$.
Hence they represent contributions from small, large 
and intermediate scales, respectively.

In addition to those terms we obtain three mixed terms that arise from
`interferences' between the different modes. Those terms depend on
$k^{-1} $ , $k^{-2}$ and $k^{-3}$.  Figure 3 compares the window
functions with and without the Sachs-Wolfe terms.  Although the
Sachs-Wolfe and Doppler effects change the window functions on very
large scales, their contribution to the derived rms $a_{lm} $ integral
for the power-spectra considered are tiny relative to the $a_{lm}$
arising from density fluctuations.  The difference in
$a_{lm}/a_{00}$ is no more than 0.1 \% 
over the harmonics range $1 \le l \le 10$.  Therefore, the
Sachs-Wolfe and Doppler effects can safely be ignored in our analysis.

\vfill\eject
%\magnification=\magstep1

\centerline{{\bf Table~1}. 
Dipole and Quadrupole moments in the X-ray Background}
\vskip 0.2 true cm
\hrule
\vskip 0.2 true cm
\tabskip=1em plus2em minus.5em
\halign to\hsize
{#\hfil&&\hfil#&\hfil#\cr
  &  Dipole $ a_{1m}/a_{00}$  & Quadrupole $a_{2m}/a_{00}$ \cr
   &       & \cr
Expected velocity dipole from COBE  &  $1.4\times 10^{-3}$&    \cr
   &       & \cr
rms velocity CDM  & $3.7 \times 10^{-3} $ & \cr
rms velocity LDCDM & $4.3 \times 10^{-3} $ & \cr
    &       & \cr
    &       & \cr
Observed HEAO1  LSS dipole & $ 4.6 \times 10^{-3}$ & \cr 
   &       & \cr
rms LSS CDM ($q=0, z_{max}=5$)   & $ 8.2 \times 10^{-3}$ &  $ 6.5 \times 10^{-3}$ \cr
rms LSS LDCDM ($q=0, z_{max}=5$)  & $ 9.3 \times 10^{-3}$ &  $ 7.2 \times 10^{-3}$ \cr
   &       & \cr
rms LSS CDM ($q=4, z_{max}=5$) &  $  4.8 \times 10^{-4}$ &  $ 3.8 \times 10^{-4}$ \cr
rms LSS LDCDM ($q=4, z_{max}=5$)& $  5.7 \times 10^{-4}$ & $  4.6 \times 10^{-4}$ \cr
   &       & \cr
rms LSS CDM ($q=4, z_{max}=3$) &  $  7.0 \times 10^{-4}$ &  $ 5.6 \times 10^{-4}$ \cr
rms LSS LDCDM ($q=4, z_{max}=3$)& $  8.3 \times 10^{-4}$ & $ 6.6 \times 10^{-4}$ \cr
   &       & \cr
rms Shot Noise (hard band) 			& $ 8.0 \times 10^{-4}$    & $ 8.0 \times 10^{-4}$ \cr
}

\smallskip\smallskip\hrule
\medskip

Comments:

(i) The predicted velocity-induced dipole is based on 
the interpretation of the COBE dipole 
being due to the motion of the Sun  at 369 km/sec
relative to the CMB.

(ii) The rms velocity is calculated in linear theory for a point,
assuming either Cold Dark Matter (CDM) or Low Density CDM (LDCDM)  
power-spectra.
The value scales like the product $\sigma_8 \Omega^{0.6}$, 
taken here to be unity.

(iii) The observed HEAO1 dipole is from Jahoda (1993), 
after correcting for Galactic emission and the 
velocity-induced dipole.

(iv) The predictions due to  large scale structure (LSS) 
assume either CDM or LDCDM
power-spectra (normalized with $b_x \sigma_8 =1$) 
in  $\Omega=1$ universe.
The perturbations are assumed
to grow like $\delta(z) \propto (1+z)^{-1}$
and the comoving emissivity to evolve like $\rho_x(z) \propto (1+z)^q$, 
given here for $q=0$ and $q=4$ 
out to redshift $z_{max}=5 $ or 3  (with $z_{min}=0$). 

(v) The shot noise was estimated for the hard band using: $N_0 \approx
2 \times 10^{-15}$ (erg/sec cm$^{-2}$)$^{3/2}$ str$^{-1}$ and $f_{m} =
3 \times 10^{-11}$ erg/sec cm$^{-2}$, above which sources were
identified (Piccinotti et al. 1982) and an observed hard-band mean
intensity (Boldt 1987) ${\bar I} = a_{00}/{\sqrt{4 \pi}}  = 5.2 \times
10^{-8}$ ergs/sec/cm$^2$/str.

\vfill\eject

\noindent{\bf FIGURE CAPTIONS} 

\bigskip

\noindent{\bf Figure 1:}  The quadrupole window functions
$|\Psi_{l=2}(k)|^2$, where 
$$ < |a_{lm}|^2 > ~\propto \int dk k^2 P(k)
|\Psi_l(k)|^2.  
$$ 
For the CMB, $|\Psi_l(k)|^2 = k^{-4}j^2_l(2ck/H_0)$
is due to the Sachs-Wolfe effect.  The quadrupole window
function of the IRAS 1.2 Jy redshift survey is based on  Fisher,
Scharf \& Lahav (1994), 
with a Gaussian radial function 
centred at 6000 km/sec with $\sigma=2000$ km/sec.
That of the XRB is given by Eq.~16 in the
text, assuming $\Omega=1$, $q - \mu=3$, $z_{min}=0 $ and $z_{max}=5$. 
The solid and
dashed lines represent $k^3P(k)$ for a standard CDM model and the
observed galaxy power spectrum (fitted by a low density CDM model)
respectively.

\bigskip

\noindent{\bf Figure 2:} XRB  rms normalized harmonics 
$a_{lm}/a_{00} b_x \sigma_8$.  
The solid and dashed lines correspond to CDM and LDCDM models 
both with $\Omega=1$, perturbation growth 
$\delta \propto (1+z)^{-1}$ and  evolution law $\rho_x(z) \propto (1+z)^q$ 
out to $z_{max} =5$ with $q=4$.
The shot noise level
normalized is  shown for two  flux limits 
above which  sources are removed from the ROSAT map.

\bigskip

\noindent{\bf Figure 3:} The Sachs-Wolfe and velocity effects on the quadrupole
window function $|\Psi_2(k)|^2$ (Eq.~A8) are shown by the dashed
lines.  The solid lines represent the same functions when these
effects are neglected. We assumed $q-\mu=3$ as in Fig.~1.  As they
only affect the largest scales, the SW and velocity effects are
significantly reduced when weighted by $k^3P(k)$ (here a low density
CDM model fitting the observed galaxy power spectrum) and their
resulting contribution to the $< |a_{lm}|^2 >$ is negligible.

\end